# Designing Building Blocks for Open-Ended Early Literacy Software

Ivan Sysoev: MIT Media Lab (corresponding author: isysoev@alum.mit.edu)
James H. Gray: MIT Media Lab
Susan Fine: Northeastern University
Deb Roy: MIT Media Lab

## Abstract

English has a convoluted relationship between its pronunciation and spelling, which obscures its phonological structure for early literacy learners. This convoluted relationship has implications for early literacy software, particularly for open-ended, child-driven designs. A tempting way to bypass this issue is to use manipulables (blocks) that are directly tied to phonemes. However, creating phoneme-based blocks leads to two design challenges: (a) how to represent phonemes visually in a child-accessible way and (b) how to account for context-dependent spelling. In the present work, we approached these challenges by developing a set of animated, onomatopoeia-based mnemonic characters, one per phoneme, that can take the shape of different graphemes. We applied the characters to a construction-based literacy app to simplify independent word-building for literacy beginners. We tested the app during a 13-week-long period with 4- to 5-year-olds in kindergarten classrooms. Children showed visible interest in the characters and properly grasped the principles of their functioning. However, the blocks were not sufficient to scaffold independent word building, leading children to rely on other scaffolding mechanisms. To test the characters' efficiency as mnemonics, we evaluated their effect on the speed and accuracy of finding phonemes on a keyboard. The results suggest that there were both children who benefitted from the characters in this task and those who performed better without them. The factors that differentiated these two categories are currently unclear. To help further research on phonetic mnemonics in literacy learning software, we are making the characters available to the research community.

## 1. Introduction

English has a nontrivial relationship between units of its spelling (letters) and its pronunciation (phonemes). For instance, letter A can represent at least six different phonemes: *[ɑ]* as in f**a**ther, *[æ]* as in c**a**t, *[eɪ]* as in n**a**me, *[ʌ]* as in dat**a**, *[ɛ]* as in c**a**re, *[ɔ]* as in **a**lso. Conversely, there are at least four ways to represent phoneme *[ʌ]*: U as in tr**u**ck, O as in c**o**me, A as in dat**a**, and TE as in lis**te**n. Some other languages, such as French and Danish, have similarly complex relationships between pronunciations and spellings [1]. This complexity obscures the phonological structure of the language, the mastery of which is crucial for early literacy learning (ibid.).

We look at this problem in the context of a particular branch of literacy technology: open-ended, child-driven learning media. These systems are related to the constructionist paradigm [3] and are motivated by the desire to facilitate not only the acquisition of "hard" literacy skills but also the formation of a positive relationship with reading and writing. Children are more likely to form a positive relationship with a subject if they have agency in their learning process and if they can meaningfully connect what they are learning to their interests and lives [2,3]. Child-driven media attempt to enable both by centering the learning process on the spelling of words chosen by children themselves and exploring the relationships between spelling and sound patterns through tinkering. Examples of such media include the apps Word Wizard [4], SpeechBlocks [5] and PictureBlocks [6]. In these systems, children were observed to exhibit initiative in building a variety of words related to their lives and interests, such as their names, favorite fictional characters, items connected to their hobbies, messages to their parents and peers, rhymes, word play and fun nonsense words [5,7]. In doing so, they showed notable agency and self-efficacy, as well as a variety of literacy-centered interactions with their peers. These factors make child-driven designs promising vehicles for early literacy learning.

Most of these designs focus on assembling words out of building blocks. Because of the complex relationship between letters and phonemes, letters might not be the optimal blocks. Indeed, to facilitate the development of phonological awareness, it is often recommended to encourage spelling "by ear", without a premature focus on orthographic correctness [8–10]. However, inconsistent, context-dependent pronunciation of letters and letter sequences (called *graphemes*) makes it hard to build words in this fashion. For instance, one of the authors observed how a child struggled to spell BEAUTIFUL in SpeechBlocks. She used B, U and T to encode the initial sounds *[b]*, *[ju]* and *[t]* and ended up with the word BUT, which is pronounced quite differently: *[b;ʌ;t]*. The child assumed that she did something wrong and disassembled her word.

These considerations make it appealing to associate building blocks with phonemes, which would make building words "by ear" relatively straightforward. To do that, however, it is necessary to represent phonemes visually so that the user can identify what is encoded by each block. If such visual representation involves letters, then it is necessary to deal with the context-dependent spelling of phonemes. We developed a set of animated onomatopoeic phoneme blocks that can turn into different letters depending on the context. We applied these blocks to an experimental child-driven early literacy app, SpeechBlocks II [7], to determine whether they encouraged independent word-building by children.

The blocks were assessed as part of a study aimed at evaluating SpeechBlocks II. The app was introduced in two kindergarten classrooms, serving 4- to 5-year-olds, for a 13-week-long period. Twenty-six children participated. To obtain a sense of the efficiency of the characters as mnemonics, at the end of the study, we provided children with a minigame asking them to locate certain sounds on either a conventional letter keyboard or a character-based keyboard. We also asked several questions to assess their understanding of the principles crucial to using the blocks. The results were mixed. While children demonstrated interest in the characters and generally well-understood their functioning, they rarely engaged in independent word building with the blocks. Instead, they relied on another autonomy-supporting mechanism, an automatic scaffolding routine. The minigame results suggest that while some children were helped by the characters in locating sounds on the keyboard, others did better with regular letters. Thus, in a child-driven context, the phoneme-based blocks and the corresponding mnemonics were of limited usefulness. However, based on related literature, it is possible that mnemonics would be more helpful to children if more systematic instruction regarding them was provided. With this in mind, we are making the characters available to the research community as a tool for studies with early literacy software.

## 2. Background

Although experienced readers recognize familiar words as a whole [11], early literacy learning is less focused on memorizing individual words and more on mastering the patterns of correspondence between sequences of letters and phonemes [12]. A significant part of early literacy learning is acquiring the phonological structure of language - how words can break down into sound units, such as syllables, onsets and rimes, and phonemes. This skill is called phonological awareness. Because the sounds of speech are blended (coarticulated), it is a nontrivial skill for a child [13], but its importance for early literacy (particularly in English) is crucial. As a reflection of the importance of phonological awareness, in the early stages of literacy development, many educators (e.g. [8,9]) recommend encouraging invented spelling: an unconventional way of spelling words (e.g., FES instead of FISH) that some children naturally develop and that is grounded in their early phonological knowledge [14,15]. Conversely, a premature focus on orthography is deemed undesirable [10].

The orthographic complexity of English presents a significant obstacle for early literacy learners, which is one of the key reasons why English learners acquire foundational word reading skills more than twice as slowly as their peers in countries with more straightforward writing systems [1]. To support the development of children's phonological knowledge, multiple authors have tried to introduce an intermediate writing system for beginners. Sandel [16] describes eight such attempts, of which the most well-known is the Initial Teaching Alphabet (Fig. 1). This alphabet has an unambiguous letter-to-phoneme correspondence, achieved by augmenting the English alphabet with several special symbols. Early experiments with the ITA showed a variety of learning gains resulting from its usage and the relative ease of transitioning to conventional orthography (ibid.). However, despite ITA's brief period of fashion, it failed to gain prominence. One possible reason for this failure is that subsequent studies showed that coloring letters to reflect the underlying phonemes could yield almost the same advantages as using the ITA [17]. However, perhaps the main reason was that the ITA required an entire new ecosystem of literacy materials that was not compatible with what was readily available in classrooms and in children's environments. At least one attempt to use a transitional spelling system also appeared in literacy learning software. Falbel [18] assigned fixed graphemes to represent each phonemes in his open-ended spelling system, Talking Blocks.

<div align="center">

# kiŋ fiʃh bʊk

</div>

Fig. 1. KING, FISH and BOOK spelled in ITA

An alternative approach to the orthographic complexity problem is to integrate some reminders about the underlying phonemes into the letters. Some works (e.g. [17,19]) use color-coding of letters for this purpose. Other designs use pictorial mnemonics, utilizing either (a) the onomatopoeic principle (evoking a sound by something that makes the sound, e.g., sneezing) or (b) the rebus principle (evoking a sound by a word that starts with this sound; e.g., using apple to represent phoneme [æ]). The onomatopoeic principle was used in educational tools/programs such as Dekodiphukan [20], Lively Letters [21], Reading Genie [22], and Leapfrog [23], while programs such as Zoo-Phonics [24], Phonics Safari [25], Itchy's Alphabet [26], and Letterland [27] explored rebus mnemonics. Studies have shown the efficiency of mnemonics for learning letter-sound correspondence and elements of phonological awareness [26-34]. Additionally, it was found that mnemonics are more effective when integrated into the shapes of letters [29, 30], and when they gradually fade as children progress [28]. It was also suggested that the stories associated with mnemonics can increase the meaningfulness of the instruction and children's motivation [27], although the evidence supporting this is currently insufficient. All of the aforementioned studies utilize rebus-based mnemonics. While we identified several studies exploring onomatopoeic mnemonics [35,36], they have significant limitations [29], and this area appears to be insufficiently researched. Furthermore, all of the studies mentioned in this paragraph focus on reading, and use designs centered on letters (e.g. showing what sounds each letter makes), not phonemes (e.g. showing what spellings each phoneme assumes).

This study explores these ideas in a new context: child-driven, expressive learning software. This context places emphasis on spelling and invokes a different lens through which to look at these ideas: whether they can be tools that can help the child engage in independent, rich construction play. Furthermore, the digital environment creates some new affordances, such as animating mnemonics to reinforce their link with phonemes. Finally, to our knowledge, our work is unique in introducing a consistent visual identity for phonemes while rendering them as multiple graphemes.

## 3. Design

## 3.1. SpeechBlocks II

An Android early literacy app called SpeechBlocks II [7], developed by the authors, was used as the base for experiments. SpeechBlocks II has an open-ended, child-driven design. The intended gameplay is similar to SpeechBlocks [5] and PictureBlocks [6] and centers around (1) tinkering with words, including the fun of building nonsense words, and (2) creating compositions out of words and images associated with them (Fig. 2). Words are built by dragging blocks from the keyboard into the construction area (*word box*) and arranging them in there. SpeechBlocks II was created to address one of the limitations of earlier expressive literacy media: the high dependence of children on adult-provided phonics guidance [7]. We explored two approaches towards this goal: (1) phoneme-based blocks and (2) automatic scaffolding systems.

Phoneme-based blocks were provided as an option. If the child did not find it convenient to work with them, then they could switch to letter mode using a dedicated button on the screen to build words in a conventional manner.

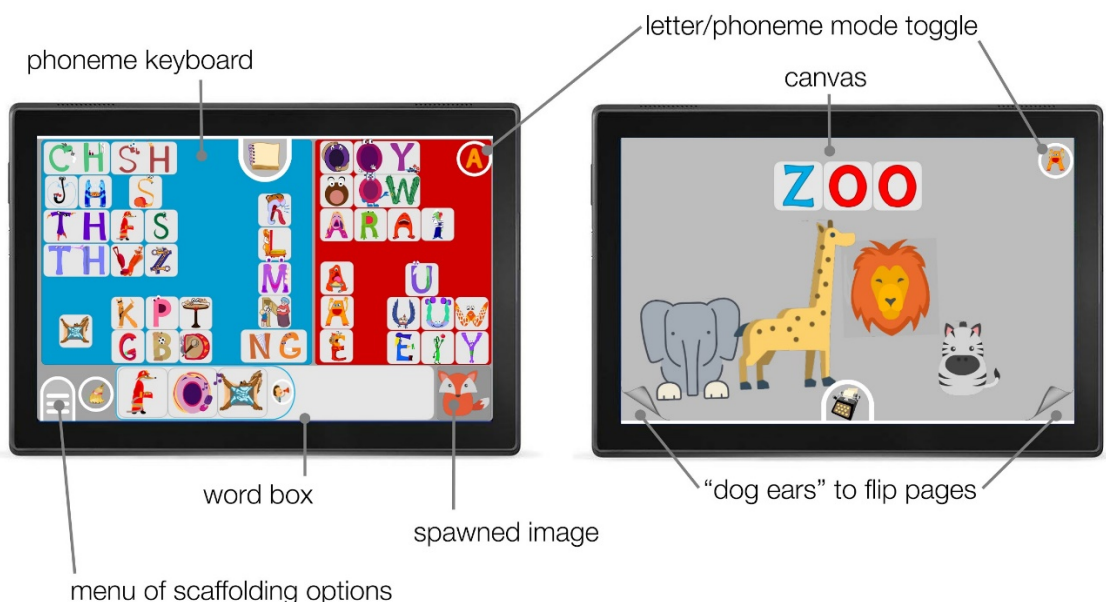

*Fig. 2. SpeechBlocks II: word construction screen and image canvas*

The automatic scaffolding system [7] received input regarding the word that the child wanted to make using several channels, such as speech recognition, semantic association network and a hierarchical word library. The system then scaffolded (i.e., guided) the process of word construction by (1) reducing the number of blocks on the keyboard, (2) enunciating the word sound-by-sound, (3) invoking cues to help locate graphemes corresponding to each sound, (4) correcting mistakes, and (5) preassembling parts of the words that it deems too difficult for the child at the moment. The present paper discusses the scaffolding system only to the extent of its connection to phoneme-based blocks.

## 3.2. Converging on a Block Design

To implement a phoneme-based block, one must choose whether its visual representation involves letters and, if yes, how to account for the context-dependent spelling of phonemes. We identified at least five categories of possible designs:

**(1) Letter-less blocks.** In this approach, phonemes are represented by visual symbols separate from letters, e.g., by mnemonic pictures.

**(2) A teaching alphabet**, such as the abovementioned ITA [16].

**(3) Fixed phoneme + grapheme blocks.** This approach was explored by Falbel [18]. It pairs each phoneme with a unique, fixed grapheme, leading to unconventional spellings, such as TOKING BLAWKS for "talking blocks".

**(4) Inverse blocks.** This is the opposite of conventional letter-based blocks: while the latter have fixed spelling and context-dependent pronunciation, inverse blocks have fixed pronunciation and context-dependent spelling. For instance, the block representing the phoneme *[ʌ]* would look like U in TR<u>U</u>CK and like TE in LIS<u>TE</u>N. This idea was previously considered, but not implemented [5].

**(5) User-controlled blocks.** In this approach, the child assembles a word out of letters but can manually adjust the pronunciation of each letter by selecting a plausible phoneme.

Theoretically speaking, approaches (1), (2) and (3) have similar strengths and weaknesses: they eschew the context dependency issue but are isolated from the rest of the literacy learning ecosystem and from environmental texts and do not allow the child to capitalize on the knowledge that comes from these streams. Approach (4) avoids this limitation, but changing spelling on its own can potentially be confusing for the users. This issue might be mitigated by creating salient design elements to mark the phoneme behind the block and thus highlight the permanent aspect of its nature. Approach (5) also avoids the issues with the previous solutions and supports both invented and conventional spellings. The challenges with this approach are (a) the sophisticated user interaction, which can lead to usability issues for young children, and (b) the ease of creating cryptic spellings, such as GHOTI for FISH[1], which can lead learners to the incorrect assumption that spellings are arbitrary, as well as confuse their peers.

We explored some of these approaches during our play-testing sessions. Since we used iterative approach, we chose which designs to try based on the observations made during previous iterations. They will be described in section 6.1. The following phoneme representations were tried:

---

[1] See https://en.wikipedia.org/wiki/Ghoti

1. Letter-less, via color-coding and corresponding visemes (i.e., the shapes that the mouth takes when pronouncing each phoneme; Fig. 3, b)
2. Letter-less, via the rebus principle (Fig. 3, c)
3. Letter-less, via the onomatopoeic principle (Fig. 4, a),
4. Via the onomatopoeic principle, with morphing into corresponding graphemes (Fig. 4, b)

The first three versions correspond to the approach (1), and the fourth – to the approach (4). Design (4) was introduced, since play-testing suggested the prominent role of logographic knowledge in early spelling.

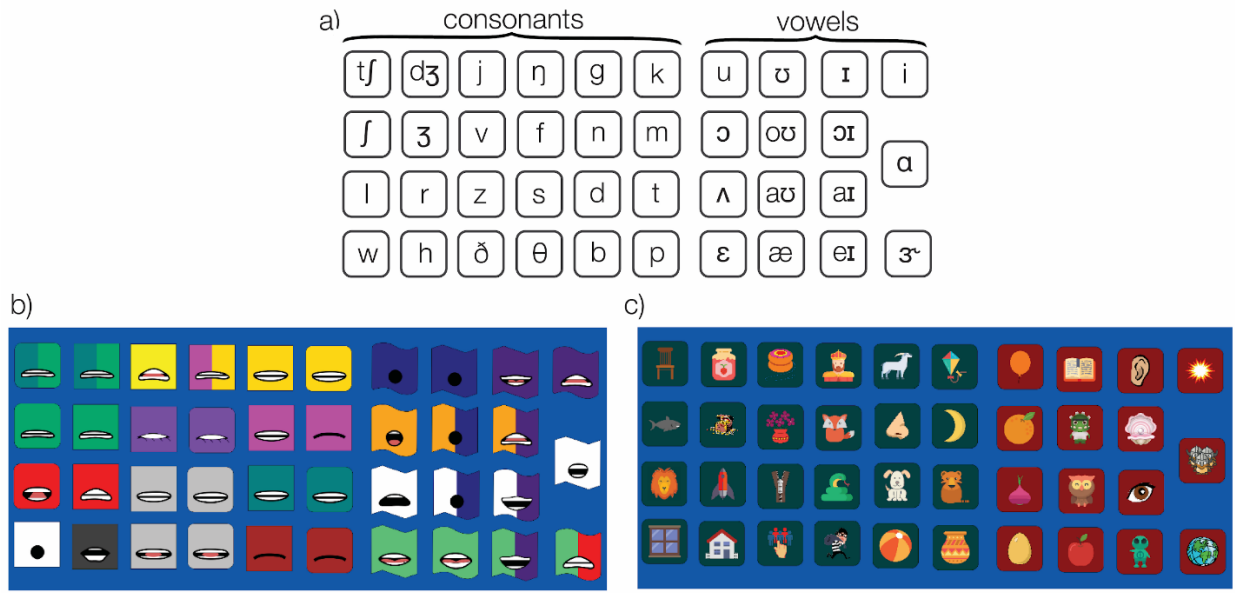

*Fig. 3. The letterless phoneme blocks we explored: (a) the keyboard layout, (b) viseme-based and (c) rebus-based blocks.*

For the latter two designs, we designed characters, or "sound creatures", performing actions that make a sound similar to the corresponding phonemes (Fig. 4). For example, a creature for phoneme *[tʃ]* sneezes: "Tch! Tch!" We used Dekodiphukan, Lively Letters, Reading Genie and Leapfrog (all referenced in section 2) as inspiration for some of the onomatopoeic actions and added a number of our own concepts. We gave each sound creature a name (e.g., "Chuck" for the *[tʃ]* creature) to reinforce their identity and to supplement the onomatopoeic principle with the rebus principle (via the first sound of the name). We also hoped that children might form parasocial connections [37] with the creatures, contributing to their memorization. Fig. 4 shows how the letter-enabled sound creatures can take the shape of different graphemes, and how the same letter can correspond to different creatures when it codes different phonemes (e.g. C in CAT and CITY). A word spelled in sound creatures makes the underlying phonemes explicit. Using the CMU Pronouncing Dictionary [38], we identified the 80 most common phoneme-grapheme combinations and created a version of a sound creature for each of them. To show the sound-producing actions more clearly and therefore to highlight the connection between the creatures and the phonemes, we animated the creatures.

These animations were made available for the research community[2]. Based on the play-testing results, the letter-enabled sound creatures were selected for deeper evaluation.

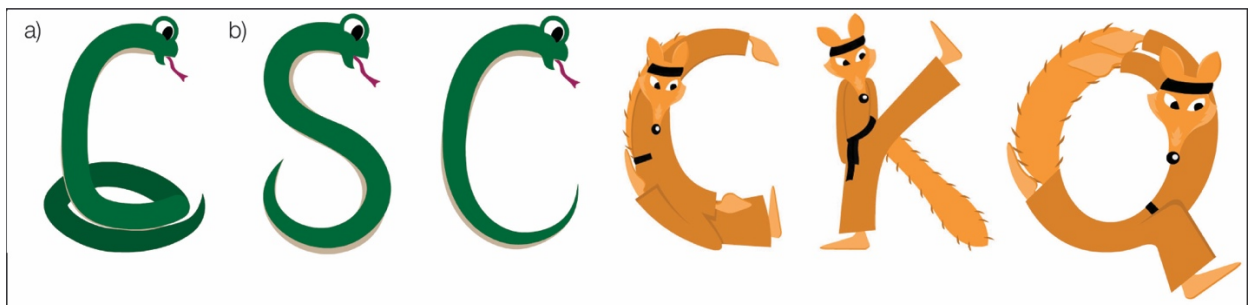

*Fig. 4. Sound creatures for [s] and [k], letter-less (a) or taking the form of different letters (b)*

The phoneme-based blocks required a dedicated keyboard. We based the layout of the keyboard on neurological research assessing the perceptual similarity of phonemes [39,40]. Consonants were clustered into three groups (fricatives, consonants, nasals and approximants), whereas the vowels were arranged according to 2D scaling of the neural response space in Mesgarani et. al. [40].

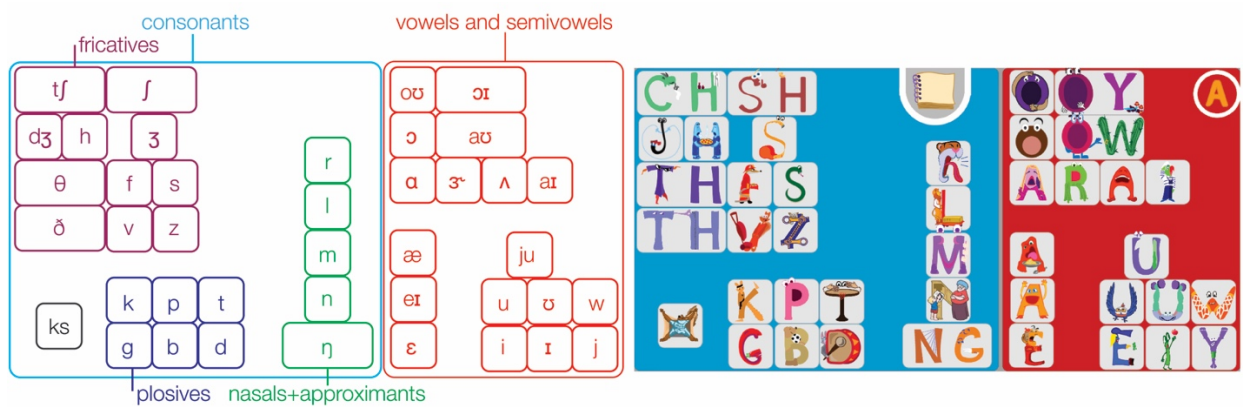

*Fig. 5. Layout and appearance of phoneme keyboard*

In section 3.1, we mentioned that SpeechBlocks II can operate in the nonscaffolded and scaffolded modes. The functioning of the blocks differs somewhat in these modes. In the nonscaffolded mode, either the pronunciations (letter mode) or spellings (phoneme mode) of blocks change depending on the context. In the scaffolded mode, the context of each block is known in advance. Therefore, it is possible to assign a fixed phoneme and grapheme to it. Switching between the letter and phoneme modes in this regime only changes whether letters or sound creatures are displayed on the blocks. In a way, the scaffolded mode allows the combination of the potential advantages of the letter and phoneme modes.

## 4. Method

---

[2] https://github.com/mitmedialab/sound-creatures

At the beginning of our work, many design choices remained unclear. We concluded that formative studies would be most useful at that stage. We conducted such studies in two phases. During the first phase, following the recommendations of Resnick et. al. [41], we performed multiple iterations of prototyping and play-testing. The goal of this was *not* to obtain rigorous or conclusive results, but to quickly zoom in on a promising design. This design, the letter-enabled sound creatures, was evaluated as part of a larger study, which focused on testing and refining various design elements of SpeechBlocks II. The study was conducted following Design-Based Research paradigm [42,43], which focuses on observation of the design in a realistic context and its gradual refinement in response to these observations.

In order to structure a design-based research work and to help derive generalizable results from it, Sandoval [44] recommends mapping the conjectures that guide the design: starting with high-level conjecture(s), indicating how it is embodied within the design, how this embodiment is supposed to generate mediating processes, and how these processes are expected to lead to learning outcomes. Our core conjecture was that bypassing the issue of orthographic complexity via special manipulables will facilitate autonomous word construction in an expressive early literacy medium. The embodiment of this conjecture was our blocks and software design.

We had in mind the following mediating process: children sounding out the desired words and building them partially by finding the identified sounds on the keyboard, partially by tinkering with pronunciations until achieving the desired result. This process is grounded in the conjectures that children would be able to perform sounding out to a sufficient degree, and that the sound creatures can be helpful in locating the needed phonemes. That in turn implies that children would understand that: (1) each creature denotes a particular sound, which is always the same; (2) the sound is associated with the creature's action; (3) the same creature can appear in multiple forms. Due to the supposed role of logographic knowledge, we also hoped that children would understand that (4) these forms depict letters or letter combinations. Since our app was child-driven, we also relied on the conjecture that children would be curious to explore the sound creatures in order to get familiar with them. Finally, we hypothesized that there could emerge a parasocial relationship between children and the creatures, which could also lead to their better memorization.

With respect to the outcomes, we surmised that the ability to effectively build words on their own would further support children's senses of agency and self-efficacy, which in turn would lead to more sustained engagement compared to earlier expressive literacy media. The resulting increase in word building, combined with the attention to the phonological structure of the words that our mediating process presupposes, could ultimately lead to better development of such skills as phonological awareness.

These conjectures determined the focus of our study. Ultimately, we were interested in observing whether and how children would use the phoneme blocks for open-ended word building, particularly in presence of such alternatives as conventional blocks and automated scaffolding. We also introduced means for examining various conjectures related to the mediating process.

## 5. Procedures and Participants

The play-testing of SpeechBlocks II was approved by the Institutional Review Board (IRB) of MIT and conducted at a children museum in the Boston area. Each session was led by one of the researchers and one assistant (a volunteer graduate student). Our participants were children aged from 4 to 7 years old. Three to six children participated in each session. Out of nearly 20 play-testing sessions with the app, 9 sessions explored the phoneme blocks. The focus of each session was, respectively: (1) building words with viseme-based blocks; (2) and (3) building words with rebus-based blocks: open-ended and children names only; (4) examining whether children recall the association between the sound creatures and phonemes after a single exposure (with initial set of 4 creatures); (5) examining whether children recognize the letter-enabled sound creatures when they take different shapes (with 6 creatures); (6) building words with sound creatures (15 creatures); (7) letter-less sound creatures (building words; 15 creatures); (8) and (9) refining arrangement of the phoneme keyboard.

The results of these sessions (elaborated in section 6.1) helped us to arrive at the "sound creatures" design. This design was tested as part of a study dedicated to the evaluation and refinement of SpeechBlocks II. The study protocol was approved by the IRBs of MIT and Northeastern University. Written consent was acquired from the families of the participants, and verbal assent was obtained from the participants themselves. The study was conducted at a public charter school in the Greater Boston area. The school was recruited through a previous working relationship with one of the researchers, who interacted with it as part of an outreach program. While we didn't gather the ethnicity and SES data about our participants, we note that the school predominantly served low- and middle-SES students of color. Two kindergarten classrooms within the school were introduced to the app. In these classrooms, there were 26 consented participants: 15 boys and 11 girls, aged 4 to 5 years old. While we don't have detailed information on the participants' early literacy skills, one bit available to us suggests that the children might have been somewhat behind on some literacy measures. Specifically, their mean PA composite score on the CTOPP-2 test was 90.4 – or about 0.64 standard deviations below the mean of the normative sample for the test.

For part of each school day, children rotated between tables ("stations") with different activities in groups of 4 to 5 and spent approximately 15 minutes on each activity. SpeechBlocks II was introduced as one of these activities. The study continued for 13 weeks, with each child interacting with the app approximately twice a week. Thus, each child accumulated approximately 6 hours of total play time with the app. At the beginning of each session, a facilitator introduced some new features of the app to the children. The participants were also assisted in case technical difficulties occurred. Phoneme blocks were introduced to children early on in the study as part of this routine. The teachers in the classrooms were not directly involved in the sessions.

We avoided video recording in the classrooms in order not to capture non-consented children. Instead, qualitative information on children's play was collected by two observers taking notes during each session. The facilitator and observer roles were performed both by the researchers and by six volunteer students from the Northeastern University's speech-language pathology program, who participated as part of their clinical practice. In order to capture the richness of open-ended play, the notes were open-ended, but we requested the observers to prioritize documenting expressions of agency and self-efficacy, evidence of children's intent (e.g. if they verbalized which words they wanted to build, or were looking for a particular letter or sound), their interactions with the sound creatures and instances of their confusion with the app. These observations helped us paint a picture of whether and how children used phoneme blocks for open-ended play, and whether sound creatures piqued their curiosity. To make sure that the observers captured the aspects of play we were interested in, we trained them using video recordings of children playing with SpeechBlocks I. The notes were analyzed inductively, by clustering interesting observations to detect emerging phenomena. In addition, detailed records of the children interactions were automatically logged by the apps. These records helped us identify how many words, and which ones, children built with or without sound creatures, and with or without automatic scaffolding.

Two additional pieces of data were collected at the end of the study. To see if the sound creatures were effective mnemonics for phonemes, we created a minigame that prompted children to press buttons corresponding to given sounds. For each sound, the players were presented either with a letter-based or creature-based keyboard. In the minigame, both keyboards followed the alphabetic layout, to minimize differences between conditions and since it was not necessary to show all the phonemes. The players were given three attempts to find each phoneme. We measured both the number of errors and the time it took the players to find the sound (if found at all). To increase the sensitivity of our statistical analysis, we collected paired data points: each child was tasked with locating each sound both on the letter and the creatures keyboards. To minimize the possibility that the children will remember the location of the key from the time they saw it in a different condition, we split the gameplay into two sessions, with each session having each task in only one condition. There was a week-long gap between the sessions, intended to make memorization less likely. We selected eight phonemes (*[r], [w], [k], [z], [m], [s], [d], [f]*) for the game, and two (*[t] and [b]*) - for children to practice with the interface of the game. The selection was based on the observation that children frequently reacted to the corresponding sound creatures during the main study, suggesting that they were relatively well-designed and memorable.

To examine our conjectures regarding children's understanding of the creatures, we also conducted structured poststudy interviews with the children. We showed the interviewees a page displaying several forms of the creature Kathy (for the phoneme [k]; see Fig. 4), selected due to being well-received by children and having several corresponding graphemes. We then asked a sequence of questions which included the following: (1) What is her name? (a basic probe for the possibility of parasocial connection), (2) (pointing to different forms of the same creature) Is this Katie? And this? Can you point to all the Katies? (to check the understanding that the creatures can take different forms), (3) (again, pointing to different Katies) What does this Katie say? And this? (to check the understanding that the represented sound is the same in each form), and (4) What do you think each Katie looks like? Does it shape resemble something you know? (to check whether children can recognize letters in the character shape). The dialogue with children was maintained by the second author, while the responses were recorded and subsequently analyzed by the first author.

## 6. Results

### 6.1. Play-Testing Results

Our play-testing sessions began with viseme-based and rebus-based block designs. With both of them, we saw that children either arranged the blocks chaotically and without a visible purpose or treated them as "letters in disguise". Consider the case of a girl named Evelyn[3]. She tried to build her name and said, "It starts with E!". She then located a block that made the sound *[i]*, corresponding to the letter's name. A facilitator asked her: "Does your name sound like ee-veline (*[i;v;ɛ;l;ɪ;n]*)?" The girl said: "No, it's Evelyn (*[ɛ;v;ɛ;l;ɪ;n]*)". The facilitator responded, "Then, you need the sound *[ɛ]*, because these are sounds, not letters". The girl objected: "No, my name starts with E (*[i]*)". The facilitator and the girl talked for some time about the distinction between letter names and sounds. Then, the child built *[i;v;ɛ;l;ɪ;n]* regardless and said, "The app reads it wrong!" The facilitator said, "This is because your name actually starts with the sound *[ɛ]*. Look!" He corrected the name and asked, "Does it sound right?" The child responded, "Yes. But my name starts with E".

A series of such observations led us to believe that the children used a mixture of two types of knowledge when spelling with the app: phonetic and visual (or logographic; corresponding to visual memory of the word as a whole). In doing so, the children were not fully aware of the intricate relationship between letters and sounds. This occasional reliance on visual knowledge would make approaches (1), (2) and (3) of the section 3 impractical for our purposes. In practice, approach (1) also appeared to not be truly detached from letters, as children continued to look for them.

It appeared that a working approach would need to account for ambiguity in the usage of each block: the speller could have used it to encode a letter or to represent a phoneme. Approach (5) allows the child to explicitly deal with such ambiguity. Alternatively, resolving this ambiguity can be delegated to the machine. In the case of SpeechBlocks II, we had built a suitable system in the form of an invented spelling interpreter, which was intended to be one of the input mechanisms for the scaffolding procedure. This system took a sequence of blocks arranged by the child and came up with possible interpretations of what they could mean. We therefore decided to rely on this automatic system. However, approach (5) remains a potentially interesting subject for further investigation.

Although we didn't assess children's literacy levels during play-testing, we noticed that children who were able to use the letterless blocks purposefully also appeared to read and sound out words with relative ease. That led us to an impression that at least some relatively advanced literacy skills are needed to operate such blocks with ease. Indeed, the rebus principle demands quick recognition

---

[3] This name is fictional to protect the child's identity but is chosen to represent the same pattern that appeared with real name

of the initial phoneme in the word, and the viseme principle demands awareness of mouth position during speech. In searching for phoneme representations that would not place such demands on pre-existing literacy skills, we proceeded to the onomatopoeic principle: the sound creatures design.

Play testing of sound creatures at the museum showed promising results. We found that, when presented with a small (4-15 items) set of creatures, the children typically were able to recall sound-creature associations after a single demonstration. This held for both letter-enabled and letterless version of the creatures. For the letter-enabled ones, we observed that children understood that a creature can take multiple forms but still sound the same.

## 6.2. Observational Results

During the 13-week study, the children generally responded to the sound creatures with interest and attention. More than half of the participants were documented as reacting to the animations in various ways. Below are the forms of engagement observed:

(1) Exclamations of amusement and delight, such as "This is hilarious!" and "Mamma mia!";

(2) Motor responses, such as jumping in their seats in pretend horror when seeing an animation of the snake creature or waving back to a waving creature;

(3) Mimicking the sounds produced by the creatures; and

(4) Commenting on the character's actions, such as responding "Ouch! His finger is bleeding!" to the animation of the *[oʊ]* creature, who touched a cactus and exclaimed "Ow!"

However, we also observed several cases in which the association with phonemes was obscured by superficial details of the creatures' design or by misinterpretations of their actions. For instance, a few children responded to karate-kicking animation for the *[k]* creature with sounds "Hiya!" and "Pfff!", which may have reflected common auditory depictions of karate in cartoons and games. These misinterpretations occurred even though the animation of the creatures produced the desired sound. Another example is associated with the phoneme *[ɛ]*, which was represented by an elderly person who, in attempt to hear better, exclaimed "Eh?" A child building ELSA from the cartoon Frozen could not believe that Elsa starts with *[ɛ]*, saying: "It is so ugly; how can it be in her name?"

Despite our hopes, the phoneme-based blocks did not lead to a sizable amount of independent word construction. Approximately 80% of the words built by the children were produced with the help of the scaffolding routine. Most of the remaining words were either random arrangements of blocks or words that children appeared to be familiar with and likely remembered visually. The reason for this result is likely the insufficient phonological awareness of most participants. On multiple occasions, we tried to assist children and found that they usually struggled to identify anything but the initial sounds of the words, unless they were clearly enunciated by us. Phoneme blocks were inadequate in a situation when phonemes could not be identified. Furthermore, even when the children managed to identify the needed sound, they sometimes struggled to locate it on the phoneme keyboard. In the case of the conventional letter keyboard, children were occasionally able to utilize their knowledge of letter sounds, combined with such a letter-finding method as an alphabet song.

The children did not exhibit difficulties in the guided mode, which was built around phoneme + grapheme blocks. However, children differed in their preference of whether to see the blocks rendered with letters or with the sound creatures (which was controlled by a dedicated button). Analysis of the data from the sound-finding minigame provides further details of why this might have been the case.

### 6.3. Interview Results

Despite the very minimal instruction at the beginning of the study, most of the children exhibited an adequate understanding of how the sound creatures functioned when answering the interview questions. When asked to point to all the images of a particular creature (Kathy for *[k]*) on the page, all of them except one pointed to different versions of the creature, showing their understanding that it remained the same character despite being morphed into different letters. Twenty-one out of 26 children recognized that all the forms of the creature produce the same sound and were able to make that sound. The 22nd child came close but produced *[k;w]* for the Q version of *[k]* creature, probably having words such as QUEEN in mind. All of the children except one recognized the letters behind the creatures, mitigating our concerns that the visual details of the creatures will prevent them from doing so. However, during the interview, we observed that two out of 26 children interpreted the action of the creature in a way that didn't connect with its sound (i.e. "makes a circle of himself, sticks a leg out").

Only three children recalled the name of the creature. The rest called it "a fox", "a ninja", or "a karate [fighter]". This finding suggests that they did not relate parasocially to the characters, which is not surprising given the lack of depth in our portrayal of the creatures. Since some phonemes only offer rare and unusual names starting with them, it might be beneficial to omit names from future designs and refer to the creatures as, say, "Mrs. *[k]*".

### 6.4. Sound-Finding Game Results

Table 1 shows that children's aggregated performance in the letter and creature modes was quite similar. However, some individual children demonstrated dramatically different performance under the two conditions. For instance, one child made 7 errors with letters and 0 with creatures, while for other children, the error tallies could be similarly skewed in the other direction. The statistical analysis presented below suggests that these differences were beyond mere chance.

Table 1. Descriptive statistics for error rates and response times in letters and creatures conditions

| Measure | Min | Q1 | Median | Mean | Q3 | Max |
|---|---|---|---|---|---|---|
| Errors per child (letter) | 0 | 1 | 3.5 | 4.69 | 7.0 | 18 |
| Errors per child (creature) | 0 | 2.25 | 3 | 4.23 | 5.75 | 10 |
| Sec. per phoneme (letter) | 0.13 | 1.73 | 3.1 | 5.97 | 7.67 | 63.37 |
| Sec. per phoneme (creature) | 0.43 | 1.77 | 3.17 | 6.76 | 8.37 | 72.9 |

In our analysis of the minigame data, we took advantage of Bayesian mixed-effects models [45]. This type of modeling offers two benefits. First, this type of modeling accounts for dependencies between data points, while simpler techniques, such as the t-test, assume that all data points are independent. This is very relevant in our case since each sound-finding task was performed by multiple children and each child performed multiple tasks. Second, Bayesian models provide posterior estimates regarding the distribution of individual idiosyncrasies, which allows us to determine whether there was a significant variability in children's response to the two conditions. We used the *brms* library for R (ibid.) to conduct the analysis.

To help explain the results, let us briefly describe mixed-effect models [46]. These models treat a given dataset as a sample from a random distribution of hypothetical datasets. These models assume that the response variable (in our case, number of mistakes, or time per phoneme) is sampled from a distribution that is parametrized by a linear combination of predictor variables. This linear combination has two types of coefficients. One type of coefficient is called *fixed effects*; these coefficients are constant among all hypothetical datasets and correspond to general trends (e.g., the effect of the treatment). The other type of coefficient is called *random effects* and represents the idiosyncrasies of participants and test items. Random effects can be further divided into *random intercepts* (which are individual additions to the bias term) and *random slopes* (which represent interactions between the participants/test items and conditions). The intercepts and slopes for each random effect are presumed to jointly come from a bivariate normal distribution. The inference algorithm uses data to estimate parameters of random effect distributions.

While modeling the time per phoneme, we used the logarithm of this value as the response variable and picked a model with a normal response distribution. We chose logarithms because the distribution of the raw times was highly skewed, while the distribution of the logarithms was close to normal. We chose to model the number of mistakes via a geometric distribution. A geometric distribution models the number of weighted coin tosses before it lands on heads.

Neither the response time nor the error number model show a significant difference between the two conditions (Table 2), with credible intervals overlapping zero even at the 90% level. However, further analysis of the data suggests that a fraction of children markedly performed better with letters and a fraction of children who markedly performed better with creatures. This can be seen by looking at the 95% credible intervals for the standard deviation of random slopes corresponding to variation in response of participants to condition. As we can see in Table 3, in the error rates model, the credible interval is well separated from zero. Combined with the main effect being small, that suggests the presence of both "letter-lovers" and "creature-lovers" (children who find sounds more accurately with letters and creatures, respectively) in the population.

Table 2. Fixed effects of the letter condition for different models (90% bounds are used to show that the credible intervals overlap with zero even at that level)

| Model | Low-90% Bound | Expectation | High-90% Bound |
|---|---|---|---|
| Times per phoneme | -0.38 | -0.14 | 0.09 |
| Error rates | -1.01 | -0.38 | 0.19 |

Table 3. Standard deviation of the distribution of random slopes for different models

| Model | Low-95% Bound | Estimate | High-95% Bound |
|---|---|---|---|
| Times per phoneme | 0.02 | 0.28 | 0.63 |
| Error rates | 0.31 | 0.94 | 1.71 |

To see how pronounced this phenomenon is, we estimated the percentages of "letter-lovers" and "creature-lovers" in the population at large (assuming that our sample was representative) for various levels of effect size (e.g., how many children find sounds at least twice as fast with creatures). This was done via the following steps: (1) use the Bayesian inference algorithm to draw samples of parameters for random effect distributions, (2) use each sample to generate a "virtual population" of hypothetical children, (3) compute the fractions of interest for the virtual population, and (4) gather the estimated fractions over all samples to form our posterior distributions of belief regarding what the true fraction is. Using this technique, we arrive at the results in Tables 4 and 5. The data suggest that, in terms of accuracy of sound finding, there is a sizable fraction of the population that would exhibit a preference for each condition. It is unclear whether the same pattern holds for response times. What differentiates "letter-lovers" from "creature-lovers" is currently unclear.

Table 4. Estimated fractions of "letter-lovers" and "creature-lovers" w.r.t. error rates.[4]

| | % of "creature-lovers" | | | % of "letter-lovers" | | |
|---|---|---|---|---|---|---|
| | Low-90% Bound | Expectation | High-90% Bound | Low-90% Bound | Expectation | High-90% Bound |
| less errors | 18.1% | 36.7% | 58.1% | 41.9% | 63.3% | 81.9% |
| >1.25x less err. | 6.3% | 22.9% | 42.7% | 26.8% | 50.3% | 70.8% |

---

[4] Here, we used 90% intervals b/c 95% intervals were too broad and uninformative due to limited data.

| | | | | | | |
|---|---|---|---|---|---|---|
| >1.5x less err. | 1.3% | 14% | 31.8% | 16.9% | 41.3% | 63.1% |
| > 2x less err. | ≈0% | 5.9% | 18.2% | 6.9% | 29.7% | 52.3% |

*Table 5. Estimated fractions of "letter-lovers" and "creature-lovers" w.r.t. time per phoneme.*

| | % of "creature-lovers" | | | % of "letter-lovers" | | |
|---|---|---|---|---|---|---|
| | Low-90% Bound | Expectation | High-90% Bound | Low-90% Bound | Expectation | High-90% Bound |
| faster | ≈0% | 30.3% | 64.3% | 35.7% | 69.7% | 100% |
| >1.25x faster | ≈0% | 12.5% | 36.5% | 1% | 38.1% | 76.1% |
| >1.5x faster | ≈0% | 6% | 22.8% | ≈0% | 18.7% | 46.7% |
| > 2x faster | ≈0% | 1.9% | 9.5% | ≈0% | 6% | 22.4% |

## 7. Discussion and Conclusion

We developed phoneme-based blocks, an alternative form of building blocks for open-ended word construction in early literacy software. Our core conjecture was that phoneme-based blocks would be more conducive for independent word construction by children by eliminating the issue of orthographic complexity. However, we saw that the introduction of such blocks did not appear to encourage self-guided word construction. It appeared that 4- to 5-year-olds struggled breaking down words into phonemes and locating blocks on the phoneme keyboard. The automatic scaffolding routine, which assisted children with both tasks, ended up being much more successful. Most of the words were constructed via the scaffolded mode, which seems to be a more promising direction of development for open-ended early literacy apps.

To visually represent phonemes on phoneme blocks, we developed a set of animated onomatopoeic mnemonics called sound creatures. We conjectured that the creatures would be able to provoke enough curiosity for children to explore them. We also surmised that children would be able to understand a few core principles of the creatures' functioning, such as that a creature could assume the form of different graphemes while still making the same sound. These conditions were necessary for the hypothesized mediating processes to unfold. These conjectures were supported by the observations and the interview data. However, a key conjecture about the creatures – that they would be helpful in locating blocks corresponding to different sounds – appears to be true only for some participants and false – for others. The results of the sound-matching game suggest that some children did not benefit from the sound creatures as mnemonics for phonemes, finding it easier to locate sounds on a keyboard when blocks were rendered with letters. Additionally, in play data, we saw that some participants were switching to the letter mode during scaffolded spelling. We did not identify the factors that contribute to children's preference for letters or creatures. One likely factor is the letter-sound knowledge. Unfortunately, it was not assessed in our present study and remains an intriguing subject for further research.

The present study complements the previous work on phoneme representations and phonetic mnemonics in several ways. First, it focuses on onomatopoeic mnemonics, which received little attention in research literature. Our observations suggest their potential advantages (i.e. lesser reliance on pre-existing phonological awareness) and disadvantages (i.e. possibility of misinterpretation). Further studies are needed to establish those with more certainty. Second, we used animations, which apparently fostered the children's interest in the creatures. This contributes to research examining the relationship between phonetic mnemonics and motivation [27]. Third, in contrast with earlier "letter-centric" designs, we assigned a distinct visual identity to each phoneme while rendering it with various graphemes. Our results suggest that children generally recognize this identity, which could be useful for such purposes as visualizing arbitrarily complex relationships between pronunciations and spellings. Fourth, we focused not on letter-sound knowledge or reading, but on self-expressive spelling. It would, therefore, be interesting to evaluate the utility of onomatopoeic and "phoneme-centric" designs, in comparison to rebus-based and "letter-centric" ones, for learning letter-sound associations. As for the spelling task, our observations suggest the importance of capitalizing on pre-existing letter-sound and logographic knowledge. Children utilize developing knowledge of these types in invented spelling. While children rarely used invented spelling in our study, it is interesting to examine (a) whether the same would be the case with older children, and (b) whether accounting for invented spelling can improve spelling scaffolding. Fifth, in contrast with almost all similar works, our study was not within the instructionist paradigm. This implies that we primarily relied on child-driven exploration for participants to get familiar with the mnemonics. That might explain why they were not effective at representing sounds for some children. In line with Brennan's recommendations [47], further consideration needs to be given to the proper combination of child's agency and externally imposed structure when introducing such mnemonics.

The present study has a number of limitations. The quantitative analysis is exploratory, and confirmatory studies are needed to firmly establish the observed patterns. The sample size is small (n=26), introducing wide ranges of estimates in the analysis. The demographic information on the sample was not collected. The indirect evidence suggests that it might not be representative of the US population, although it might be an interesting case of underserved population. Some potentially relevant measures, such as the children's letter-sound knowledge, were also not obtained. The study was conducted by the same researchers who developed the assets, introducing potential researcher bias. The observation notes that we used to collect qualitative data and interview results could have led to observer bias and imprecisions (e.g. observers missing some details while documenting other events). The mini-game and the interview were administered with limited subsets of the creatures (8 and 1, respectively), making it possible for the observed results to be specific to these designs. This might also apply to the entire set of 80 creatures. While we invested our best effort in designing the creatures, we didn't have an opportunity to refine each of them via iterative play-testing. Thus, the optimality of the mnemonics and animations is not certain. Additionally, the rapid prototyping results obtained in the first phase of the study are by their nature far from conclusive, and other regions of the design space might well be worth exploring. Despite these limitations, we believe that this study presents an interesting exploration case and is a useful starting point for future work.

One contribution of the present work is the set of onomatopoeic mnemonics itself. These designs can be useful for such studies evaluation of onomatopoeic and phoneme-centered mnemonics and experiments with explicit presentation of grapheme-phoneme mapping within words. We hope that the research community will find the creatures useful assets.

## Acknowledgments

We would like to acknowledge (1) Sneha Makini, who provided multiple valuable assets from her app PictureBlocks to assist development of SpeechBlocks II; (2) Sneha Makini, Nazmus Saquib and Manuj Dhariwal – for assisting with play-testing; (3) Allie Shoff, Christine Schlaug, Kathryn Falvey, Anne Mathew, Mackenzie Russell, and Lauren Brinker – for their vital help in facilitating the sessions and collecting observations; (4) Allan and Daniel Gelman, Jesso Wang and R. Ryan Hayes – for animating a large fraction of the sound creatures; Abraham Tena and Lingxi Li – for contributing several additional animations; (5) Elena Bodrova, Catherine Snow and Anneli Hershman – for ideas and feedback; (6) Hae Won Park, Safinah Ali and anonymous reviewers – for providing invaluable feedback on the paper; (7) Twitter – for providing financial support for the MIT team; (8) teachers and staff at the museum and the school – for welcoming us and kindly accommodating our various requests.

This research did not receive any specific grant from funding agencies in the public, commercial, or not-for-profit sectors.